\documentstyle
[12pt]
{article}
\begin{document}
\newcommand{\ed}{\end{document}}
\newcommand{\pr}{\prime}
\newcommand{\ppr}{\prime\prime}
\newcommand{\cE}{{\cal E}}
\newcommand{\vphi}{{\varphi}}
\newcommand{\oO}{O(k^{-1})}
\newcommand{\be}{\begin{equation}}
\newcommand{\ee}{\end{equation}}
\newcommand{\barr}{\begin{array}}
\newcommand{\earr}{\end{array}}
\newcommand{\bea}{\begin{eqnarray}}
\newcommand{\eea}{\end{eqnarray}}
\newcommand{\pa}{\partial}
\newcommand{\xx}{\hbox{}^*_*}

\title{Coset conformal field theories with
abelian isotropy groups.}
\author{A.V.Bratchikov\thanks{bratchikov@kubstu.ru}\\ Kuban State
Technological University,\\2 Moskovskaya Street, Krasnodar, 350072,
Russia.} \date{November,1998} \maketitle
\begin{abstract}
Conformal field theories based on $g/u(1)^d$ coset constructions
where $g$ is a
reductive algebra are studied.It is shown that the theories are
equivalent to constrained WZNW models for $g.$ Generators of extended
symmetry algebras and primary fields are constructed.
\end{abstract}

In a recent paper \cite {B} currents,
primary fields and generators of disctete symmetry groups of
$g/u(1)^d, d=1,\ldots rank\,g,$ coset conformal field theories \cite
{BH,GKO} for simple Lie algebras $g$ were constructed.The construction
is based on operator quantization of the WZNW theory for $g$ with the
constrained affine Lie algebra $\hat u(1)^d$ associated with $u(1)^d.$
In this article we consider the $g/u(1)^d$ coset theory where $g$ is a
direct sum of simple and abelian Lie algebras.Using a version of the
generalized canonical quantization method \cite {BFF, B} we show that
this theory is equivalent to a constrained WZNW theory and find
parafermionic currents and primary fields.They are expressed in terms
of the initial fields and free bosons.The theory has a discrete
symmetry group which is generated by an abelian subalgebra of $g.$

Studies of such theories were initiated in \cite {FZ} where
$su(2)/u(1)$ parafermionic algebra was constructed and it was shown
that the corresponding theory describes a critical system.

Let us consider the WZNW theory \cite {W,KZ} for
$g = g_{1}+ g_{2}+\ldots+ g_{M},$
where $g_{\mu}$ are simple Lie algebras for $1\leq \mu \leq M^\prime$
and $g_{\mu}=u(1)^{r_{\mu}}$ for $M^\prime+1 \leq \mu \leq M.$We treat
only the golomorphic part of the theory. Let $k_\mu$ be the central
charge of the affine Lie algebra $\hat g_{\mu}$ associated with
$g_{\mu}.$

The affine Lie algebra $\hat g=\hat g_{1}+\hat
g_{2}+\ldots+\hat g_{M}$ is generated by
elements of the form
\bea \label{E}
E(z)=\sum_{\mu^\prime=1}^{M^\prime}\sum_{\alpha\in\Delta_{\mu^\prime}}
b^{\alpha}_{\mu^\prime}E_{\mu^\prime}^\alpha(z)+
\sum_{\mu^{\ppr}=M^\prime+1}^M\sum_{s=1}^{r_{\mu^{\ppr}}}
b^{s}_{\mu^{\ppr}}H_{\mu^{\ppr}}^s(z)
\eea
where $\Delta_{\mu^\pr}$ is the set of the roots and
$E_{\mu^\pr}^\alpha(z)$ are step generators of $g_{\mu^\pr}$,
$H_{\mu^{\ppr}}^s(z)$ are generators of $\hat u(1)^{r_{\mu^{\ppr}}},$
$b=(b_{\mu^\pr}^\alpha,b_{\mu^{\ppr}}^s)$ are numbers.

Let $g_{\mu^\prime}$ be simply-laced.In this case the currents
$E_{\mu^\pr}^\alpha(z),H_{\mu^{\ppr}}^s(z)$ and the energy-momentum
tensors $L_\mu(z)$ associated with $\hat g_\mu$ via Sugawara
construction can be expressed in terms of the bosonic fields
\begin{equation}\label{boson} \varphi_\mu^{sj}(z)=x_\mu^{sj}-ia_{\mu
0}^{sj}logz+i\sum_{n\not =0}{a_{\mu n}^{sj}\over n}z^{-n},
\end{equation} where
$s=1,\ldots r_\mu,r_{\mu^\pr}
\equiv rank\, g_{\mu^\pr},j=1,\ldots k_\mu,\>\mu=1,\ldots M$ and
nonvanishing commutators of the modes are given by
\begin{equation} [x_\mu^{s j},a_{\mu 0}^{sj}]=i,\qquad
[a_{\mu n}^{sj},a_{\mu(-n)}^{sj}]=n.  \end{equation} The bosonic
construction for $E_{\mu^\pr}^\alpha(z)$ and $H_{\mu^{\ppr}}^s(z)$
reads \cite{FK} \bea E_{\mu^\pr}^\alpha(z)
=\sum^{k_{\mu^\pr}}_{j=1}:e^{i{\alpha}\cdot\varphi_{\mu^\pr}^j(z)}:
c_{\mu^\pr\alpha}^j,
\qquad
H_{\mu^{\ppr}}^s(z)
=\sum^{k_{\mu^{\ppr}}}_{j=1}i\partial_z\varphi_{\mu^{\ppr}}^{sj}(z).
\eea
where :: denotes normal ordering with respect to the
modes of the bosons,${\alpha}^2=2$ and $c_{\mu^\pr \alpha}^j$ is a
cocycle operator.The bosonic construction for $L_{\mu^\pr}$ was
obtained in \cite {DHS}. Let $ H_{\mu^\pr}^s(z),1 \leq s \leq
r_{\mu^\pr} $ be the currents of $\hat g_{\mu^\pr}$ which are defined
by \bea H_{\mu^{\pr}}^s(z)
=\sum^{k_{\mu^{\pr}}}_{j=1}i\partial_z\varphi_{\mu^{\pr}}^{sj}(z).
\eea

Let  $G^\lambda(z),
\lambda=(\lambda_\mu^s),s=1,\ldots r_\mu,\mu=1,\ldots M,$
be the primary field of $\hat g$ and the Virasoro algebra generated by
$L^g=\sum^M_{\mu=1}L_\mu$ which satisfies the equations \bea H_{\mu
n}^sG^{\lambda}&=&L_{\mu n}G^{\lambda}=0,\qquad \mbox {for}\quad
n>0,\nonumber \\ H_{\mu 0}^sG^{\lambda}&=&\lambda^{s}_\mu
G^{\lambda},
\\
L^g_oG^{\lambda}&=&{\Delta(\lambda)}G^{\lambda},\qquad
L^g_{-1}G^{\lambda}=\partial_{z}G^{\lambda},\nonumber
\eea
where $H_{\mu n}^s$ and $L_{\mu n}$ are modes of the corresponding
fields: $H_{\mu}^s(z)=\sum_nH_{\mu n}^sz^{-n-1}\,$
$ L_{\mu}(z)=\sum_n L_{\mu n}z^{-n-2}.$

Let us consider the $\hat u^d(1)_k,k=\sum^M_{\mu=1}k_\mu,$ subalgebra of
$\hat g$ generated by the currents
\bea H^A(z)=\sum^M_{\mu=1}H_\mu^A(z) \eea
where $A=1,\ldots,d,$ $d\leq r_\mu,\mu=1,\ldots,M.$
We shall consider the WZNW theory subject to the constraints
\begin{equation}\label{constr}H^{A}(z)\approx0,
\end{equation}
This equation can be rewritten in the form
\begin{equation}H^{A}_n\approx0,
\end{equation}
where $H^{A}_n=\sum_{\mu=1}^MH_{\mu n}^A.$

To quantize the system we put into correspondence with the
constraints the bosonic fields
\begin{equation}\label{boson}
\phi^{A}(z)=q^A-i\tilde a^A_0logz+i\sum_{n\not
=0}{\tilde a_{n}^{Aj}\over n}z^{-n},
\end{equation} where
nonvanishing commutators of the modes are given by
\begin{equation}\label{gamma}
[q^{A},\tilde a^{A}_0]=ik,\qquad [\tilde a_n^{A},\tilde a_{-n}^{A}]=nk.
\end{equation}

     According to a version of the generalized canonical quantization
method \cite {BFF,B} we replace the constraints (\ref{constr}) by the
effective abelian constraints \begin{equation}{\tilde
H}^{A}(z)=H^{A}(z)+\partial \phi^A(z)\approx0.  \end{equation} It is
easy to see that \begin{equation}{\tilde H}^{A}(z){\tilde H}^{B}(w)=0.
\end{equation}

The current $E(z)$ (\ref{E}) is replaced  by
\bea
\label{cE} {\cE}(z)=\sum_{\mu^\pr=1}^M\sum_{\alpha\in\Delta_{\mu^\pr}}
b^{\alpha}_{\mu^\pr}{\cE}_{\mu^\pr}^\alpha(z)+\sum_{\mu^{\ppr}
=M^\prime+1}^M\sum_{s=1}^{r_{\mu^{\ppr}}}
b^{s}_{\mu^{\ppr}}{\cal H}_{\mu^{\ppr}}^s(z), \eea
\bea
\cE_{\mu^\pr}^\alpha(z)=E_{\mu^\pr}^\alpha(z):e^{\frac{1}{k}
\tilde\alpha\cdot\phi(z)}:
,\eea
\bea
{\cal H}_{\mu^{\ppr}}^s(z)=H_{\mu^{\ppr}}^s(z)+
\frac{k_{\mu^{\ppr}}}{k}\sum_{A=1}^d\delta^{sA}\partial\phi^A(z)
,\eea where $\tilde\alpha=(\alpha^A).$ The new current commute with
${\tilde H}^A(z)$ \bea {\tilde H}^A(z)\cE(w)=0.
\eea
The $g/u(1)^d$ energy-momentum tensor $K=L^g-L^{u(1)^d},$ where
\bea
L^{u(1)^d}=\frac 1 {2k}\sum_{A=1}^d:\left({H^A}\right)^2:,
\eea
also commutes with ${\tilde H}^A(z)$ \cite{GKO}
\bea {\tilde H}^A(z)K(w)=0.
\eea

The operator product expansion of $K$ with  ${\cE}$ is
given by
\bea
K(z)\cE(w)&=&\frac{1}{(z-w)^2}\biggl(\cE(w)-
\frac{1}{2k}\sum_{{\mu^\pr}=1}^{M^\prime}
\sum_{\alpha\in\Delta_{\mu^\pr}}\tilde\alpha^2
b^{\alpha}_{\mu^\pr}{\cE}_{\mu^\pr}^\alpha(w)\biggr)\nonumber \\
&+&\frac{1}{z-w}\partial_w{\cE}(w)-\frac{1}{k}\delta \cE (b,{\tilde
H}^A),\label{ke1} \eea
where \bea \delta \cE (b,{\tilde
H}^A)&=& \frac{1}{(z-w)^2}\sum_{\mu^{\ppr}=M^\pr+1}^M
\sum_{A=1}^d k_{\mu^{\ppr}}b^{A}_{\mu^{\ppr}}{\tilde H}^{A}(w)
\nonumber \\
&+&\frac{1}{(z-w)}
\biggl(\sum_{\mu^\pr=1}^{M^\pr}
\sum_{\alpha\in\Delta_{\mu^\pr}}
b^\alpha_{\mu^\pr}
\tilde\alpha\cdot\tilde
H(w){\cE}^\alpha(w)\nonumber \\
&+&\sum_{\mu^{\ppr}=M^\pr+1}^M\sum_{A=1}^d
k_{\mu^{\ppr}}
b^{A}_{\mu^{\ppr}}
\partial_w{\tilde H}^{A}(w)\biggr)\nonumber
.\label{ke2}
\eea
Note that $\delta \cE (b,0)=0.$

If $b^{\alpha}_{\mu^\pr}$ satisfy the equations
\bea
\tilde\alpha^2 b^{\alpha}_{\mu^\pr}=\nu^2 b^{\alpha}_{\mu^\pr},
\eea
where $\nu^2$ is a constant,then
eq.(\ref{ke1}) can be rewritten in the form \bea \label{conf}
K(z)\cE(w)=\frac{\Delta\cE(w)}{(z-w)^2}+
+\frac{\partial_w{\cE}(w)}{z-w}-\delta \cE (b,{\tilde H}^A),
\eea
where
\bea
\Delta=1-\frac {1} {2k}\nu^2.\nonumber
\eea

Let $\Omega$ be the algebra
gen\-er\-ated by ${\cE}(z)$ and $K(z).$
Let us define the following set of the fields
\begin{equation}\Upsilon
=\left(U\in\Omega\vert\, U\vert_{{\tilde H}^A=0} =0 \right).
\end{equation}
Using Wick theorem one
can check that for arbitrary $U\in \Upsilon$ and $X\in \Omega$
\begin{equation}UX\in \Upsilon, \qquad  XU \in \Upsilon
.\end{equation} Hence $\Upsilon$ is an ideal of $\Omega$
and the quotient $\Omega/\Upsilon$ is an algebra.We shall
denote by $\{X(z)\}$ the coset represented by the field $X(z).$

The coset field $\{K(z)\}$ satisfies
the $g/u(1)^d$ Virasoro algebra:
\begin{equation}\{K(z)\}\{K(w)\}={c_{g/u(1)^d}\over
2(z-w)^4}+{2\{K(w)\}\over(z-w)^2}
+{{\partial_w \{K(w)\}\over{z-w}}},\end{equation}
where $c_{g/u(1)^d}=c_{g}-d.$

It follows from eq. (\ref{conf}) that $\{{\cal E}(z)\}$
is a primary field
of the $g/u(1)^d$ Virasoro algebra:
\begin{eqnarray}\label{cospr}
\{K(z)\}\{{\cE}(w)\}={{\Delta\{{\cE}(w)\}}\over(z-w)^2}
+\frac{\partial_w\{{\cE}(w)\}}{z-w}.
\end{eqnarray}

The theory has the discrete symmetry group which is generated
by $H^s_{\mu 0},s=1,\ldots r_\mu,\mu=1,\ldots M.$

The initial primary field $G^\lambda$ is replaced by
\begin{equation}{\cal G}^\lambda(z)
={G}^\lambda(z): e^{\frac 1 {k}\tilde\lambda\cdot\phi(z)}:,
\end{equation} where
$\tilde\lambda^A=\sum^M_{\mu=1}
\lambda^A_\mu.$
This field satisfies the following
equations \bea\label{pri} {\tilde H}_{n}^A{\cal
G}^{\lambda}=0,\quad \mbox{for} \quad n\ge 0,\qquad
K_n{\cal G}^{\lambda}=0,\quad \mbox{for}\quad n>0,\nonumber \\
K_o{\cal G}^{\lambda}={\Delta(\lambda,\tilde\lambda}){\cal
G}^{\lambda},\qquad
K_{-1}{\cal G}^{\lambda}=\partial_{z}{\cal G}^{\lambda}- \frac 1 {k}
\tilde\lambda\cdot{\tilde H}_{-1}{\cal G}^{\lambda},
\eea
where \begin{equation}\label{andim}
{\Delta(\lambda,\tilde
\lambda)}=\Delta{(\lambda)}-\frac 1 {2k}\tilde\lambda^2.
\end{equation}

Let $T$ be the space which is obtained by applying the
currents of $\Omega$ repeatedly to the field
${\cal G}^\lambda.$ Let \begin{equation} V=\left({\cal G}\in
T\vert \,{\cal G}\vert_{{\tilde H}^A=0}=0\right).  \end{equation} The
space $V$ is an invariant subspace with respect to the algebra
$\Omega.$ It follows from eq.(\ref {pri}) that
$\{{\cal G}^\lambda\}\in T/V$ which is represented by
${\cal G}^\lambda$ is a primary field of the coset Virasoro
algebra.

The space $T/V$ can be decomposed according to the transformation
properties under the algebra generated
by $H^s_{\mu 0}, \mu=1,\ldots M, s=1,\ldots r_\mu .$.The field
$\{{\cal G}^{\lambda}\}$ belongs to the subspace with the
charges $\lambda^s_\mu$:
\begin{equation}
H_{\mu o}^s\{{\cal G}^\lambda\}=\lambda^s_{\mu}\{{\cal
G}^\lambda\}.
\end{equation}

The theory can be formulated in terms of the initial operators
$x^{sj}_\mu$ and $a_{\mu n}^{sj}$.To show this we first note that the
coset $\{{\cal E}(z)\}$ can be represented by the field \bea \tilde
{\cal E}(z)={\cal E}(z)\vert_{\tilde a^A_n=ia_n^A}.  \eea
It follows from (\ref{cospr}) that $\tilde {\cal E}$ satisfies the
equation \begin{eqnarray}\label{prim} K(z)\tilde {\cal E}(w)
={{\Delta_\alpha\tilde {\cal E}(w)}\over(z-w)^2}
+\frac{\partial_w\tilde {\cal E}(w)} {{z-w}} .  \end{eqnarray}
$\tilde {\cal E}$ still depends on the auxiliary operators $q^A.$
However these operators commute with $x^{sj}_\mu,a_{\mu
n}^{sj}$ and hence one can set $q^A=0.$ The field ${\cal E}_0=\tilde
{\cal E}\vert_{q^A=0}$ satisfies eq.(\ref {prim}) as well as $\tilde
{\cal E}.$

These results can be generalized to non-simply-laced algebras using
the vertex operator representation of the associated affine KM algebras
\cite {GNOS,BT}.


\begin{thebibliography}{99}
\bibitem{B} A.V.Bratchikov,\,$g/u(1)^d$ parafermions from
constrained WZNW theories. hep-th/9712143.  \bibitem{BH}K.Bardakci and
M.B.Halpern,{\it Phys.Rev.} {\bf D3}(1971)2493.\\ M.B.Halpern,{\it
Phys.Rev.} {\bf D4}(1971)2398.  \bibitem{GKO} P.Goddard,A.Kent and
D.Olive,{\it Phys.Lett.} {\bf 152B} (1985) 88.
\bibitem {BFF}I.A.Batalin,E.S.Fradkin and T.Fradkina,{\it Nucl.Phys.}
{\bf B332} (1990) 723.  \bibitem{FZ} V.A.Fateev and
A.B.Zamolodchikov,{\it Sov.Phys.JETP} {\bf 82} (1985) 215.
\bibitem{W}E.Witten,{\it Comm.Math.Phys.} {\bf 92} (1984) 455.
\bibitem{KZ} V.Knizhnik and A.B.Zamolodchikov,{\it Nucl.Phys.} {\bf
B247} (1984) 83.  \bibitem {FK} I.B.Frenkel and V.G.Kac,{\it
Inv.Math.}{\bf 62}(1980)23.  \bibitem {S} G.Segal,{\it
Commun.Math.Phys.} {\bf 80} (1981) 301.  \bibitem {DHS}
G.V.Dunne,I.G.Halliday and P.Suranyi,{\it Nucl.Phys.} {\bf B325} (1989)
526.
\bibitem {GNOS} P.Goddard,W.Nahm,D.Olive and A.Schwimmer, {\it
Commun.Math.Phys.} {\bf 107} (1986) 179.
\bibitem {BT} D.Bernard and J.Thierry-Mieg, {\it
Commun.Math.Phys.} {\bf 111} (1987) 181.

\end{thebibliography}
\end{document}